\documentclass[%
 reprint,
 amsmath,amssymb,
 aps,
 longbibliography,
]{revtex4-1}

\usepackage{graphicx}
\usepackage{dcolumn}
\usepackage{bm}
\newcommand{\e}{\hat{\bm e}}

\begin{document}

\preprint{APS/123-QED}

\title{Rolling pepper shaker on a slope}

\author{Mizuki Ono and Hirofumi Wada}
\affiliation{%
 $^{1}$ Department of Physical Sciences, Ritsumeikan University, Kusatsu, Shiga 525-8577, Japan\\
}%


\date{\today}

\begin{abstract}
Although the rolling of a solid object is a mundane phenomenon in our daily life, its movement can be surprisingly complex and physically rich, particularly when the solid object has certain internal degrees of freedom, such as a half-filled plastic bottle of water. The translational and rotational motion of such an object couple in a highly nontrivial manner, often leading to seemingly unpredictable trajectories. 
We use a combination of experimental and theoretical approaches to analyze the rolling behaviors of a rigid cylinder that is partially filled with granular media, rolling on an inclined plane. 
We experimentally find a wide variety of rolling behaviors, including damped oscillation leading to a stop, meandering with avalanches coming to a stop, in addition to the stationary rolling and rolling with a constant acceleration. 
We address the occurrence of substantial slip during rolling, in contrast to what is often assumed. 
We classify the rolling behavior into three distinct phases and establish a phase diagram. 
We theoretically explain the transition between stopping and rolling and rationalize the phase boundary based on the rigid-body mechanics combined with the statics of granular media. 
Our study addresses the curiosity to understand the everyday phenomena and has significant implications for a wide range of physical applications from powder manufacturing technologies to robotics. 
\end{abstract}

\pacs{Valid PACS appear here}
\maketitle


\section{Introduction}
As can be easily imagined, when a plastic bottle of water on a flat floor is kicked, it shows a jerky motion in which translation and rotation appear to be coupled in a curious way~\cite{Jackson-Am.J.Phys.-1996, Bourges-Am.J.Phys.-2018}. 
This is due to the inertia of the water inside the container, which causes it to resist being displaced when the force is applied to the container, subsequently inducing relative motions between the plastic bottle and the confined water~\cite{Jackson-Am.J.Phys.-1996, Bourges-Am.J.Phys.-2018}. 
This can be directly confirmed by observing that the same bottle of water smoothly rolls down a steep incline with the free surface of water staying almost horizontal. 
Because the gravitational acceleration acts equally on the bottle and water, they roll down seemingly without any relative displacement~\cite{Supekar-arXiv-2014} 
(but see also Refs.~\cite{Jackson-Am.J.Phys.-1996, Bourges-Am.J.Phys.-2018} regarding the non trivial physics behind this simple phenomenon). 

\begin{figure}[b]
\begin{center}
 \includegraphics[width=0.86\linewidth]{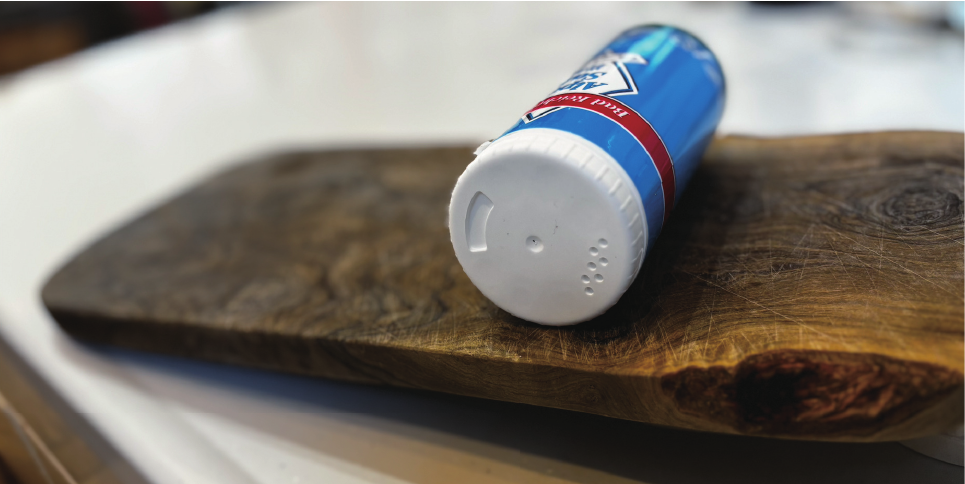}
\caption{Simple experiment that can be performed in a kitchen: partially filled salt (or pepper) shaker rolling down a slightly tilted breadboard.
If the amount of salt in the shaker is appropriate, the shaker may stop along the slope.}
\label{fig:kitchen}
\end{center}
\end{figure}

On the same inclined plane, the rolling behavior of the bottle differs distinctively when it contains granular materials instead of water~\cite{Jaeger-Rev.Mod.Phys.-1996,Dube-Chem.Eng.Sci.-2013,Chou-PowTech.-2012,Lee-PowTech.-2013,Chand-PhysicaA-2012,Wang-Phys.Rev.Res.-2024,Mellmann-PowTech.-2001,Oba-Phys.Rev.Fluids-2026,Orpe-Phys.Rev.E-2001}. 
This has been impressively demonstrated by Tadashi Tokieda in his public lecture~\cite{Tokieda-Lecture-2018}, for the case of a rigid cylindrical container filled with an appropriate amount of granular matter. 
For a suitable range of inclination angle, the rigid cylinder shows rather irregular, jerky motions, including even stopping at the middle of the slope. 
One can easily demonstrate this phenomenon by rolling a partially filled salt or pepper shaker on a breadboard (Fig.~\ref{fig:kitchen}). 
Intuitively, a physical origin of this seemingly mysterious movement may be attributable to static and dynamic friction in granular media~\cite{Ward-PowTech.-2012,Albert-Phys.Rev.E-1997,Lu-Particuology-2014,Third-PowTech.-2010}. 
However, several basic questions naturally arise. 
What is the critical inclination angle at which a cylinder starts rolling? 
How does the critical angle depend on the amount of granular media confined? 
What types of movements appear when the cylinder rolls down an inclination?
How does a rigid-body movement affect granular flow behavior, such as the apparent repose angle?
To answer these questions, a systematic, quantitative investigation is required. 

We address this issue by combining experimental and analytical approaches. 
We investigate the rolling behavior of a rigid cylindrical container containing granular materials (glass or aluminum beads) on an inclined plane. 
We observe a variety of rigid body motions as well as the behavior of granular materials inside the rigid body under a wide range of filling rates and slope angles. 
Based on the experiment, we broadly classify the rolling behaviors into three distinctive regimes and construct a phase diagram. 
We then rationalize our findings using a simple analytical theory that combines rigid-body mechanics with granular statics.
Using a high-speed camera, we break down the simple assumption of the no-slip rolling due to the granular inertia.
We also quantify the dynamic angle of repose characterizing the granular shear layer formed during stationary rolling.

The remainder of this paper is organized as follows.
In Sec.~\ref{sec:expt}, we describe the experimental setup.
In Sec.~\ref{sec:results}, we show the rolling behaviors experimentally observed, some of which we analytically rationalize.
In Sec.~\ref{sec:summary}, we summarize the findings and provide future perspectives.
We extend our detailed analytical arguments in Appendices~\ref{appendix_A} and \ref{appendix_B}.

\section{Experimental setup}~\label{sec:expt}
\begin{figure}[b]
\begin{center}
 \includegraphics[width=0.97\linewidth]{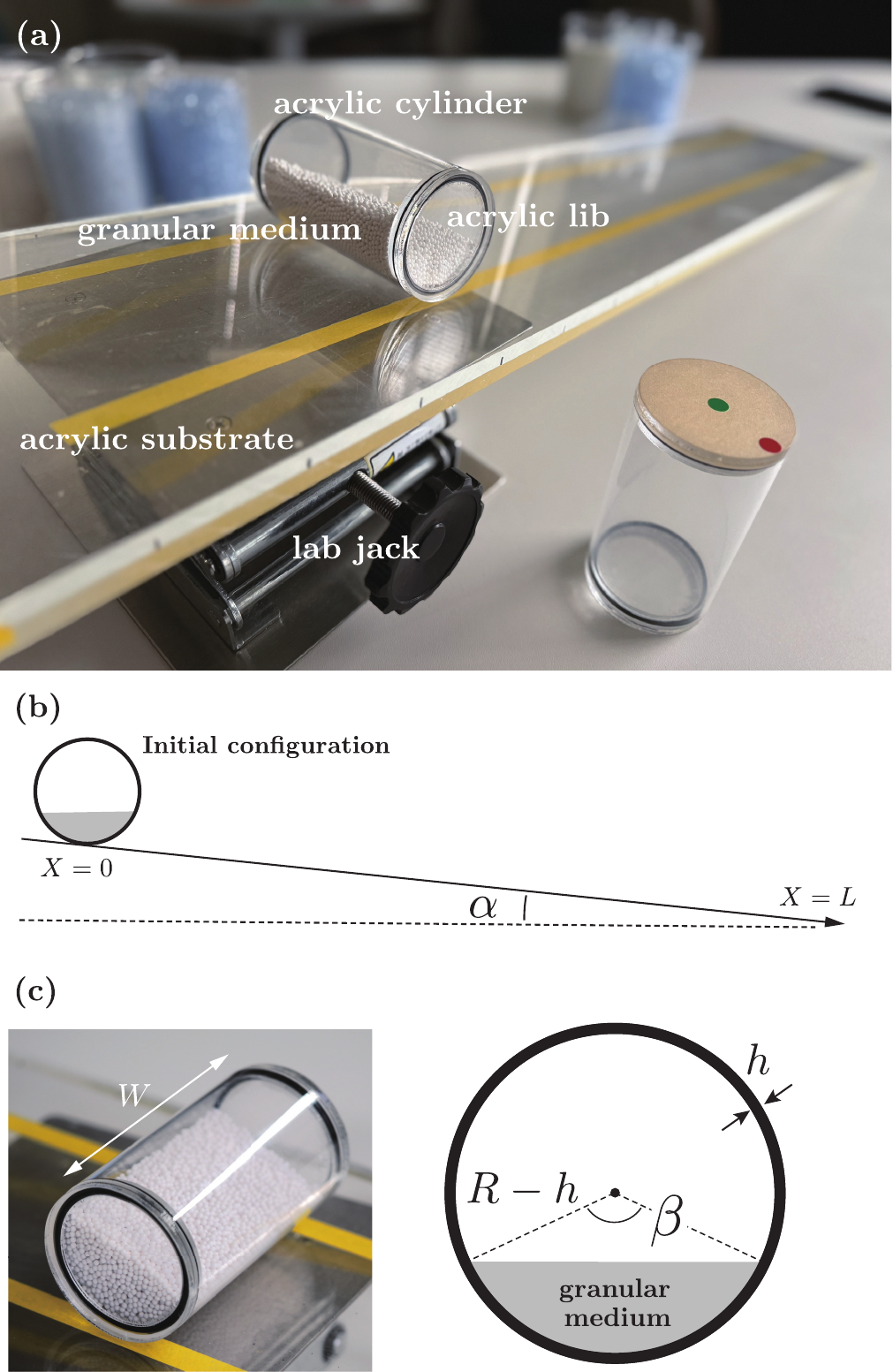}
\caption{Overview of the experimental apparatus. (a) Acrylic container on an inclined acrylic substrate, with granular medium inside.
The wood-colored lid visible in the lower right corner, with red and green markings, was used for long-term measurements of translational and rotational velocities.
(b) Geometrical parameters of the experimental system. 
A container with a horizontal layer of granular material starting to roll on a slope from rest.
(c) Close view of the container and granular medium inside. (d) Definition of the geometric parameters used to quantify the granular mass and distribution.}
\label{fig:experiment}
\end{center}
\end{figure}

Figure.~\ref{fig:experiment} shows the schematic of our experimental apparatus consisting of a flat acrylic plate of a thickness of 16 mm and length of 100 cm, on which an acrylic cylindrical vessel containing granular materials can roll down. 
The lab jack allows us to vary the slope, which we measure by the angle $\alpha$ defined in Fig.~\ref{fig:experiment}. 
We also put another acrylic box [not shown in Fig.~\ref{fig:experiment} (a)] to support the slope at its middle to minimize possible minor plastic deflection of the inclined substrate that might occur over time. 
We custom-made an acrylic cylindrical container of inner radius $R-h=27.9$ mm, outer radius $R=30$ mm, where $h=2.1$ mm is the thickness, and axial length $W=100$ mm, with the removable lids on both ends.
The mass of the main hollow cylinder and lid were, respectively, $m_s=43$ g and $m_l=33$ g, amounting to the total mass $M_c=m_s+2m_l=109$ g. 
We used alumina or glass beads of diameter $d=2$ mm as our granular materials.
The amount of grains inside the container was accurately quantified by the filling rate $f$ defined as 
\begin{equation}
 f = \frac{m}{\rho_g V_0},
 \label{eq:def-f}
\end{equation}
where $m$ represents the mass of the granular medium, $V_0=\pi (R-h)^2 W$ is the inner volume of the container, and $\rho_g$ is the effective mass density of the grains including the vacancies.
We separately measured the mass at the maximal packing, $\rho_g V_0$, and determined $\rho_g\approx 2.19\,\mathrm{g/cm^3}$  for aluminum beads and $\rho_g\approx 1.53\,\mathrm{g/cm^3}$ for glass beads.  
Taking $V_g$ as the volume occupied by the grains, we have $f=V_g/V_0$ from Eq.~(\ref{eq:def-f}); the filling rate $f$ is thus essentially the volume fraction of the grains. 

In the experiment, we gently placed an object by hand on the slope, 10 cm below the top edge, to ensure that the surface of the granular layer stayed horizontal. 
Thus, the maximum distance the cylinder could roll was $L=90$ cm.
The rolling behavior was captured by the digital camera (Pentax K-70) set five meters from the system, from which various physical quantities, such as the instantaneous position, velocity, and angular velocity, were obtained through image analysis using ImageJ. 
We also used a high-speed camera (Ditect, Japan) to capture granular flows.
For each parameter set specified by $(\alpha, f)$, we performed (at least) ten independent experiments, the average of which is shown in the following section, with appropriate error bars if needed. 
The effective length of the substrate ($< 90$ cm) limited the long-term observation of the rolling dynamics. 
To minimize transient behaviors, the study focused on the shallow slope regime, typically given by $\alpha \leq 12^{\circ}$.

\section{Results}~\label{sec:results}
\subsection{Rigid body rolling}
\begin{figure}
\begin{center}
 \includegraphics[width=0.99\linewidth]{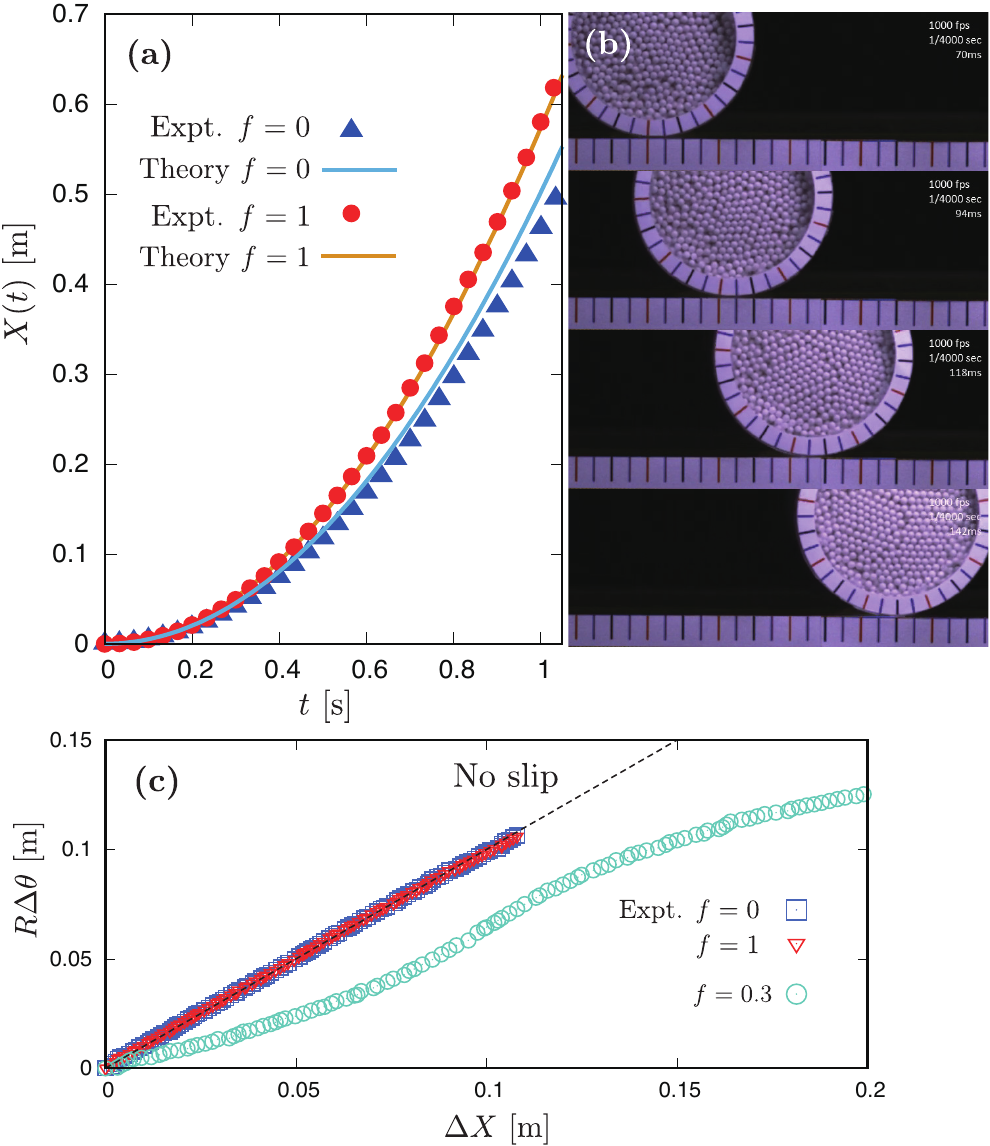}
\caption{Rolling dynamics of empty ($f=0$) and fully packed ($f=1$) containers, which are expected to behave as rigid bodies.
(a) Travel distance along the slope, $X(t)$, as a function of time $t$, compared with the simple analytical prediction given in Eq.~(\ref{eq:a-rigid}), for $\alpha=10^{\circ}$.
(b) Sequence of high-speed images of the rolling cylinder with $f=1$; time proceeds from top to bottom, with the interval between consecutive images being 24 ms.
See also Supplemental Movie S1~\cite{SM}.
(c) Non-slip condition: Travel distance, $\Delta X$, plotted as a function of rotated arclength of cylinder, $R\Delta \theta$, during the images shown in (b) for $f=0$ and $f=1$.
The dashed line suggests the non-slip condition $\Delta X= R\Delta \theta$. The data for $f=0.3$ shows a significant slip, which is discussed later. }
\label{fig:rigid-body}
\end{center}
\end{figure}

To set the stage, we first investigated the cases of $f=0$ (empty) and $f=1$ (maximally packed), both of which were expected to show simple rigid body motions with the no-slip condition. 
For a uniform cylinder of (outer) radius $R$, the acceleration $a$ along the slope is given by 
\begin{eqnarray}
  a &=& \frac{g\sin\alpha}{1+I_f/(M_fR^2)},
  \label{eq:a-rigid}
\end{eqnarray}
where $g$ is the gravitational acceleration, and $M_f$ and $I_f$ are, respectively, the total mass and the moment of inertia about the long axis of the object of filling rate $f$.
For the empty container, $f=0$, we have $I_0=(m_s+m_l)R^2$ and $M_0=m_s+2m_l$, where $m_s$ and $m_l$ are the mass of the hollow cylindrical part and lid, respectively.
For the fully packed container, $f=1$, we find $I_1=I_0+m_g(R-h)^2/2$ and $M_1=M_0+m_g$, where $m_g$ is the granular mass at its maximal packing.
Because Eq.~(\ref{eq:a-rigid}) predicts the constant acceleration, we expect $X(t)=(a/2)t^2$.
In Fig.~\ref{fig:rigid-body} (a), we compare the analytical predictions with the experimental data for $\alpha=10^{\circ}$.
An excellent quantitative agreement between the experiment and theory is observed, particularly for the full cylinder $f=1$.
Images capturedhigh-speedgh speed camera shown in Fig.~\ref{fig:rigid-body} (b) (see also Supplemental Movies S1 and S2~\cite{SM}) reveal the no-slip rolling for $f=0$ and 1, as quantitatively confirmed 
in Fig.~\ref{fig:rigid-body} (c). 
The theory predicts that the no-slip condition holds for $\tan\alpha<\mu(1+MR^2/I)$, which is satisfied here, where $\mu=0.3-0.4$ are the static friction coefficients between the acrylic surfaces.
The small disagreement observed in the $f=0$ data at the later time stage could be due to the simplified estimation of $I_0$ as the mass distribution of
the empty container is more complicated than that of the fully packed object of $f=1$.

\subsection{Classification of rolling behaviors}
Having established the accuracy of the experimental system, we investigate a variety of rolling behaviors with varying filling rates $f$ for the different slope angles, $\alpha=2, 5$ and $10^{\circ}$. 
Figure~\ref{fig:experiment} shows the displacement $X(t)$ rescaled by $2\pi R$ as a function of time $t$ rescaled by the microscopic characteristic time $\sqrt{d/g} \simeq 0.045$ s, where $d= 2$ mm is the typical diameter of either glass or aluminum beads. 
For the shallowest slope case $\alpha=2^{\circ}$, the cylinder shows damped oscillations and finally stops motion for most of $f$ studied. 
For slightly more inclined cases $\alpha=5^{\circ}$, the cylinder again stops for $f$ of approximately 0.5; however, for $f=0.1$ and 0.7, it rolls with seemingly constant velocity for $\alpha=6.5^{\circ}$.
As discussed later in more detail, this regime is close to the critical condition, and a variety of complex behaviors are observed. 
We define this domain as the “unstable” phase, in addition to the two other more definite phases. 
Finally, for the largest slope case $\alpha=10^{\circ}$ studied, the cylinder rolls all the way down the incline for most of $f$ studied. 

\begin{figure}
\begin{center}
 \includegraphics[width=0.89\linewidth]{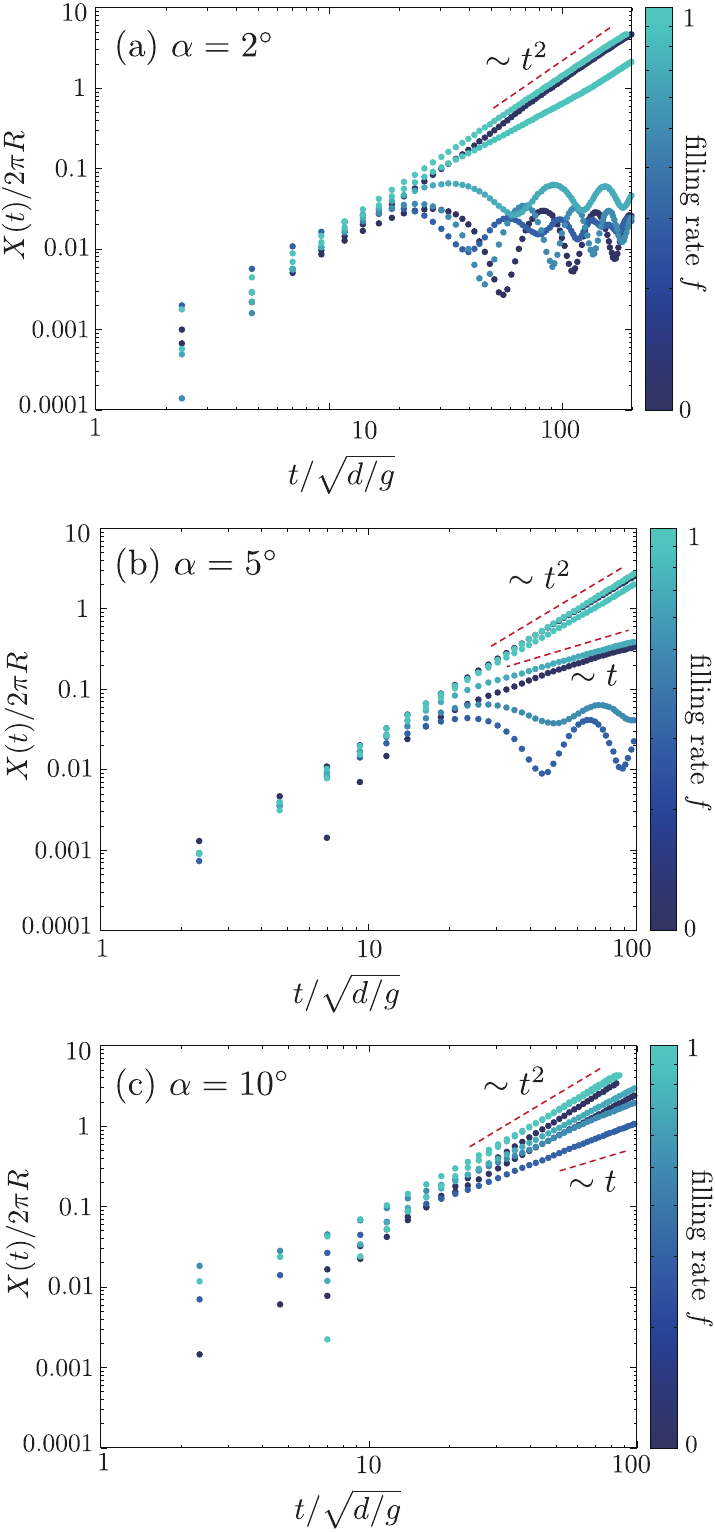}
\caption{Travel distance of the geometric center C, $X(t)$ (rescaled by $2\pi R$), along an inclined plane of various tilt angle $\alpha$ for increasing filling rates $0< f <1$ indicated with a colormap. 
(a) $\alpha = 2^{\circ}$, (b) $\alpha=5^{\circ}$, and (c) $\alpha=10^{\circ}$. Time $t$ is rescaled by the microscopic time unit $\sqrt{d/g}$.}
\label{fig:trajectory}
\end{center}
\end{figure}

Thus, for a given travel distance $L$, the observed rolling behaviors can be broadly classified into the three distinct phases, that is, (S): the damped oscillation eventually leading to the stop close to the initial poisiton, (CA): the constant acceleration translation, that is, $X(t)\sim t^2$, and (US) the unstable phase between (S) and (CA), which covers a variety of complex behaviors including the seemingly constant velocity rolling, meandering, slip with granular avalanches, or even stop, often with the axis of rotation off from the initial direction halfway down the slope.

\subsection{Phase diagram}
From the data shown in Fig.~\ref{fig:trajectory}, we can generally assume that $X(t)\sim t^{\gamma}$, where $1\leq \gamma \leq 2$.
However, the determination of the precise value of the exponent $\gamma$ is limited by the maximum distance the container can roll down, $L$.
Naively, one might expect that the work done by gravity on the entire object per unit time dissipates via the frictional convective granular flow, leading to $X(t)\sim t$, or $\gamma \simeq 1$, as observed in Fig.~\ref{fig:trajectory} (b) and (c). 
However, this argument is not convincing enough.  
Because all the friction is driven by the internal forces only (in the absence of any slip),  the center of mass of the entire object should not experience any external force that can resist the gravitational acceleration body force. 
The seemingly linear behavior, $X(t)\sim t$, may rather involve external resistive forces due to partial slip.
Therefore, in the dominant regime of rolling dynamics, we expect that $X(t)\sim t^2$, that is, (CA), is the terminal state~\cite{Supekar-arXiv-2014}.
However, a sufficiently long travel distance is required to precisely determine the value of the exponent $\gamma$, practically unattainable in the current experimental setup. 
Given this fundamental limitation, we use the above-mentioned three-phase classification to further analyze the observed rolling behaviors. 

\begin{figure}
\begin{center}
 \includegraphics[width=0.99\linewidth]{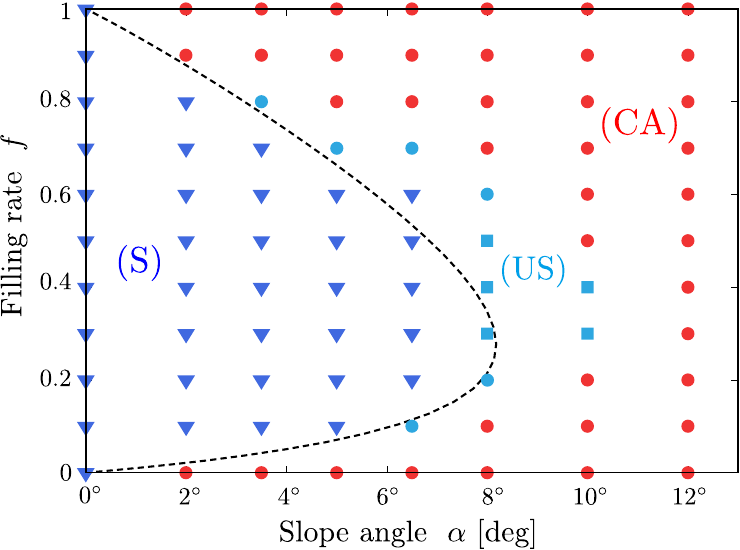}
\caption{Experimental phase diagram on the $(\alpha,f)$ plane classifying the distinct behaviors of a granular-filled rigid container. 
The blue triangles represent the stable stop configuration (S), while the red circles correspond to the constant-acceleration rolling motion (CA).
Behavior that cannot be clearly classified under either of these two phases is indicated by a cyan colored symbol, which we classify under the unstable phase (US).
US consists of a variety of motions, including a seemingly constant velocity rolling, meandering, slipping with avalanches, and even stopping, all of which are not quite reproducible.
The dashed line is the theoretical prediction given by Eqs~(\ref{eq:cond-1}) and (\ref{eq:b}), with all the parameters determined experimentally.}
\label{fig:diagram}
\end{center}
\end{figure}

For gentle slopes with $\alpha \lesssim 5^{\circ}$, the cylindrical container stops halfway down the incline in the range of the filling rates. 
To determine the boundary between  (S) and (CA), we conducted many experiments, systematically changing the filling rate $f$ and the slope angle $\alpha$, and classified the observed motions into the three phases defined above. 
By compiling extensive experimental data, we established the phase diagram on the $(\alpha,f)$ plane in Fig.~\ref{fig:diagram}, in which the blue symbols represent (S), while the red region corresponds to (CA), with the remainder of the region corresponding to (US) shown by the plum colored symbols. 
The diagram shows that the stop phase (S) is limited within the very gentle slope angles $\alpha$, being maximal at $\alpha\lesssim 8^{\circ}$ for the intermediate filling rates $f=0.3-0.5$.
We rationalize this observation with a simple analytical argument in Sec.~\ref{sec:static}.

\subsection{Static mechanical balance}~\label{sec:static}
\begin{figure}
\begin{center}
 \includegraphics[width=0.97\linewidth]{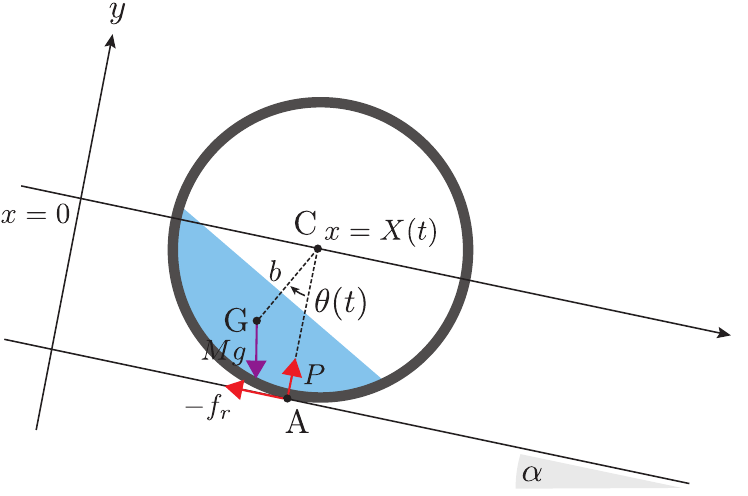}
\caption{Schmatic of the model problem analyzed and definitions of the geometrical parameters.}
\label{fig:model}
\end{center}
\end{figure}

Suppose that the entire object remains at rest on an inclined plane of angle $\alpha$, and take the tilted $xy$ Cartesian coordinates as shown in Fig.~\ref{fig:model}.
We denote A as the contact point between the cylinder and the incline, and define the rotational angle from the $y$ axis as $\theta$. 
(See also Appendix~\ref{appendix_A} for the detailed analysis.)
The condition of the moment balance is given by 
\begin{equation}
 \frac{dU(\theta)}{d\theta} = - Mg R \sin\alpha+Mg b\sin(\theta+\alpha) = 0,
 \label{eq:moment}
\end{equation}
where $U(\theta)$ is the effective potential given in Eq.~(\ref{eq:U}) and the parameter $b$ represents the distance from the geometric center C to the center of mass G of the entire object. 
Whenever a solution $\theta_0$ exists for Eq.~(\ref{eq:moment}), the equilibrium configuration is shown to be mechanically stable, that is, $d^2U(\theta)/d\theta^2 |_{\theta_0} >0$.
For further theoretical details, see Appendix.~\ref{appendix_A}.

In addition to the rigid body statics, the mechanical equilibrium requires that the granular configuration is also statically stable.
In the laboratory frame, the granular slope angle is given by $\theta_0+\alpha$, which must not exceed the angle of repose, $\phi_c$, that is, $\theta_0+\alpha <\phi_c$, 
from which we can immediately obtain the condition 
\begin{eqnarray}
 \alpha &\leq& \alpha_c(\beta) = \sin^{-1}\left[\frac{b(\beta)}{R}\sin\phi_c\right].
 \label{eq:cond-1}
\end{eqnarray}
Here, the parameter $b$, the distance between G and C, can be given by the function of the angle $\beta$ defined in Fig.~\ref{fig:experiment} (d).
Assuming that the granular medium is uniformly distributed, with the effective mass density $\rho_g$, its center of mass (of the medium alone) is expressed parametrically as
\begin{align}
 a(\beta) &= \frac{2(R-h)}{3\pi f(\beta)}\sin^3 \frac{\beta}{2},
 \label{eq:a}
\end{align}
where the filling rate $f(\beta)$ changes from 0 to 1 as $\beta$ increases from 0 to $2\pi$, given by
\begin{eqnarray}
 f(\beta) &=& \frac{1}{2\pi}\left(\beta-\sin\beta\right).
 \label{eq:f}
\end{eqnarray}
For details, see Appendix~\ref{appendix_B}.
Combining these, we obtain
\begin{eqnarray}
 b(\beta) &=& \frac{2(R-h)}{3\pi}\frac{\tau}{1+\tau f(\beta)}\sin^3 \frac{\beta}{2},
 \label{eq:b}
\end{eqnarray}
where we introduce the non-dimensional parameter $\tau=\pi (R-h)^2\rho_g/M_c \approx 4.33$.
Equations~(\ref{eq:cond-1}) and (\ref{eq:b}) determine the phase boundary curve, $[f(\beta), \alpha_c(\beta)]$, on the diagram shown in Fig.~\ref{fig:diagram}.
Plugging the experimental values, $R=30.0$ mm, $h=2.1$ mm, $\rho_a=1.17$ g/cm$^3$, $\rho_g=2.19$ g/cm$^3$, and $\phi_c=25^{\circ}$ into Eqs~(\ref{eq:cond-1}) and (\ref{eq:b}),
as shown in Fig.~\ref{fig:diagram},  the analytical prediction and experimental results are in good agreement without any adjustable parameters.

\subsection{Damped harmonic oscillation}
To better understand the small-amplitude dynamics, we consider the damped harmonic oscillation. See Fig.~\ref{fig:trajectory} (a).
From the analysis detailed in Appendix~\ref{appendix_A}, the eigenfrequency $\Omega$ is predicted in Eq.~(\ref{eq:omega}) as
\begin{eqnarray}
 \Omega &=& \sqrt{\frac{Mg\sqrt{b^2-R^2\sin^2\alpha}}{I_{C}+M(R^2-2bR\cos\theta_0)}},
 \label{eq:omega-main}
\end{eqnarray}
where Eq.~(\ref{eq:theta_0}) is used to eliminate $\cos(\theta_0+\alpha)$ in Eq.~(\ref{eq:omega}) and $\cos\theta_0$ is given in Eq.~(\ref{eq:cos}).
The frequencies are measured directly from the trajectory data shown in Fig.~\ref{fig:trajectory} (a) and (b), which are plotted as a function of $f$ and compared with the prediction Eq.~(\ref{eq:omega-main}).
Again, an excellent quantitative agreement is observed between the results of the experiment and theory without any adjustable parameters, thus confirming the overall validity of our physics argument.
\begin{figure}
\begin{center}
 \includegraphics[width=0.90\linewidth]{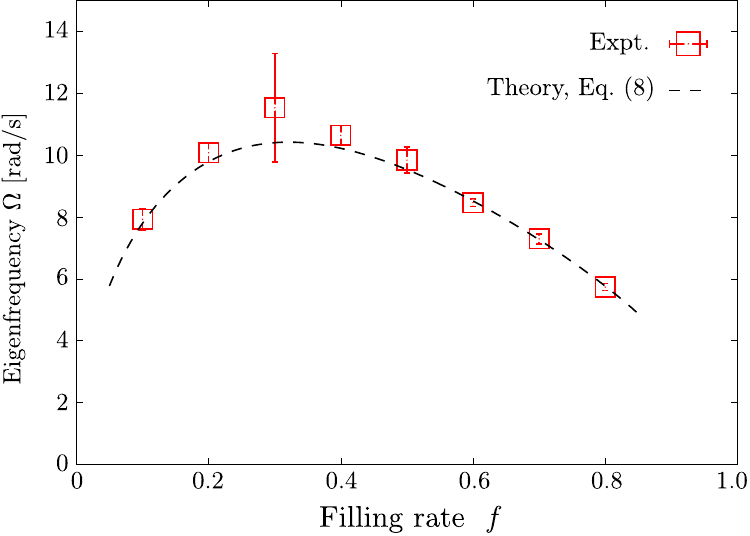}
\caption{Experimentally determined eigenfrequencies $\Omega$ for $\alpha=2^{\circ}$ [Fig.~\ref{fig:trajectory} (a)], plotted as a function of the filling rate $f$.
The dashed line represents the analytical prediction, Eq.~(\ref{eq:omega-main}), for the parameter set used in the experiments.}
\label{fig:omega-comp}
\end{center}
\end{figure}

\subsection{Instability leading to slip}
\begin{figure}[bt]
\begin{center}
 \includegraphics[width=0.67\linewidth]{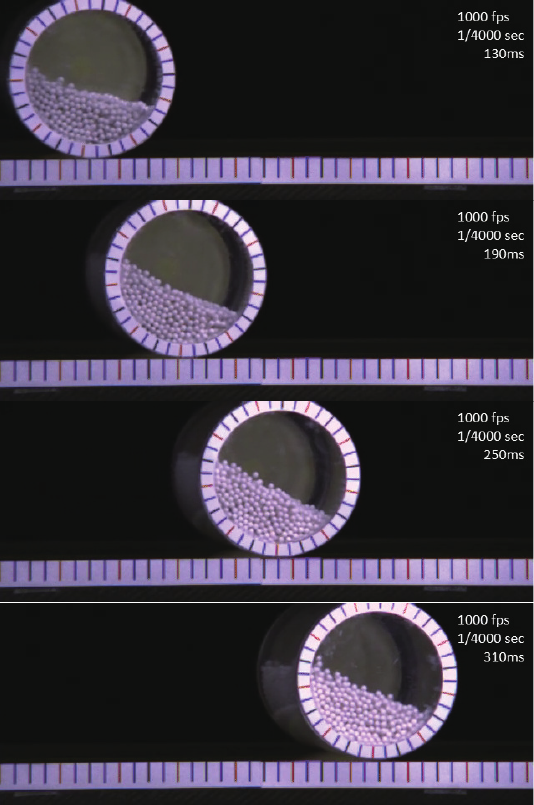}
\caption{Images captured by a high-speed camera for $f=0.3$ and $\alpha=12^{\circ}$ with alumina beads, showing a substantial slip during rolling.
Time proceeds from top to bottom, with a 60-ms time interval between consecutive images.
See also Supplemental Movie S3~\cite{SM}.}
\label{fig:slip}
\end{center}
\end{figure}

As briefly mentioned above, the behavior of the container becomes much more complicated near the boundary line in Fig.~\ref{fig:diagram}.  
Remarkably, the container may slip significantly when the stop configuration becomes unstable in the region immediately outside the (S) domain.
In Fig.~\ref{fig:slip}, we show the high-speed images for $\alpha=12^{\circ}$ and $f=0.3$, revealing the substantial slip during the rolling. 
See also the Supplemental Movie S3~\cite{SM}. 
This slip is also quantified in Fig.~\ref{fig:rigid-body} (c) as the green open symbols. 
While the simple non-slip criterion, $\tan\alpha<\mu(1+MR^2/I)$, is still satisfied for the parameter set considered here, the inertia of the convective granular flow may be responsible for the observed substantial slip.
While it is important to more precisely address how this slip physically occurs, it is beyond the scope of this study and is left for future research. 
It is worth noting that internal degrees of freedom can easily complicate slippage in the rolling problem.

\begin{figure}[bt]
\begin{center}
 \includegraphics[width=0.99\linewidth]{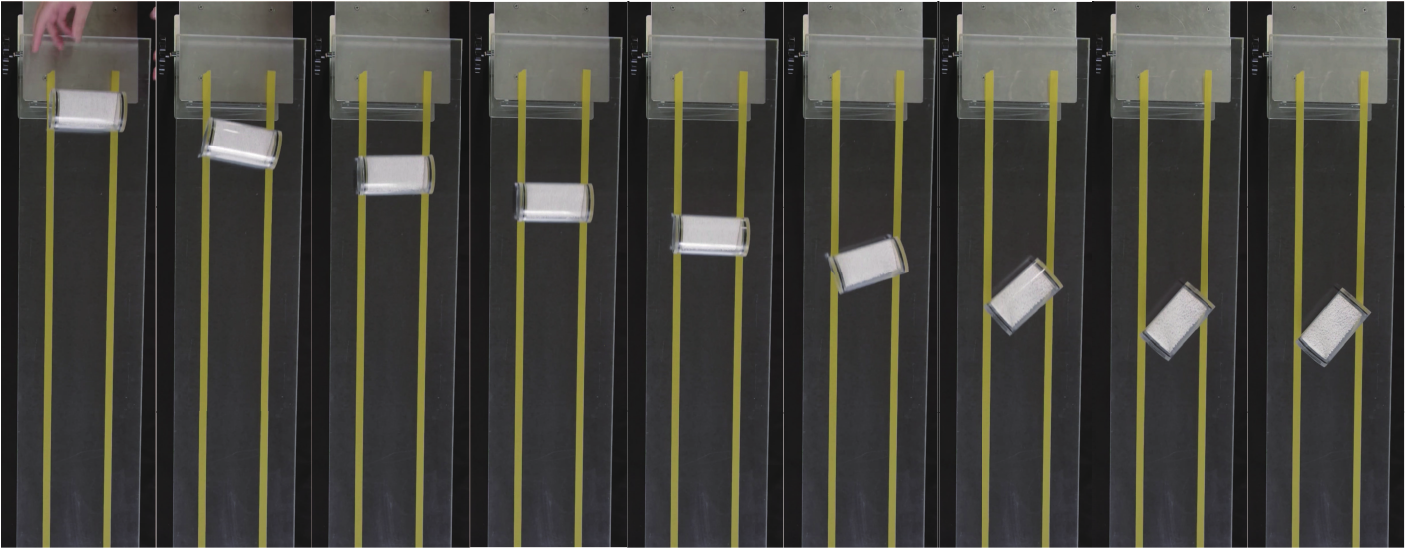}
\caption{Partially grain-filled cylinder meandering on its way down the slope and eventually coming to a stop, observed for $f=0.4$ and $\alpha=8^{\circ}$. 
See also Supplemental Movie S4~\cite{SM}.}
\label{fig:non-planar}
\end{center}
\end{figure}

Under US, other physical factors also affect the resulting container behavior. 
When the cylinder starts rolling down the slope, it can be slow and irregular due to the intermittent granular flow.
The granular configuration uniformity along the cylindrical axis was occasionally broken, turning the rotational axis of the cylinder off from the $z$ axis~\cite{Qian-PNAS-2025}.
This symmetry breaking makes the rolling behavior non-planar, leading to the meandering trajectory. See Fig.~\ref{fig:non-planar} and Supplemental Movie S4.
This type of phenomenon often results in the stopping in the middle of the slope, with the cylindrical axis being considerably off its initial direction. 
However, again, the behavior was also found sensitive to the initial configurations as well as other experimental factors that were not sufficiently precision-controlled. 

\subsection{Steady rolling and granular dynamics} 
\begin{figure}
\begin{center}
 \includegraphics[width=0.99\linewidth]{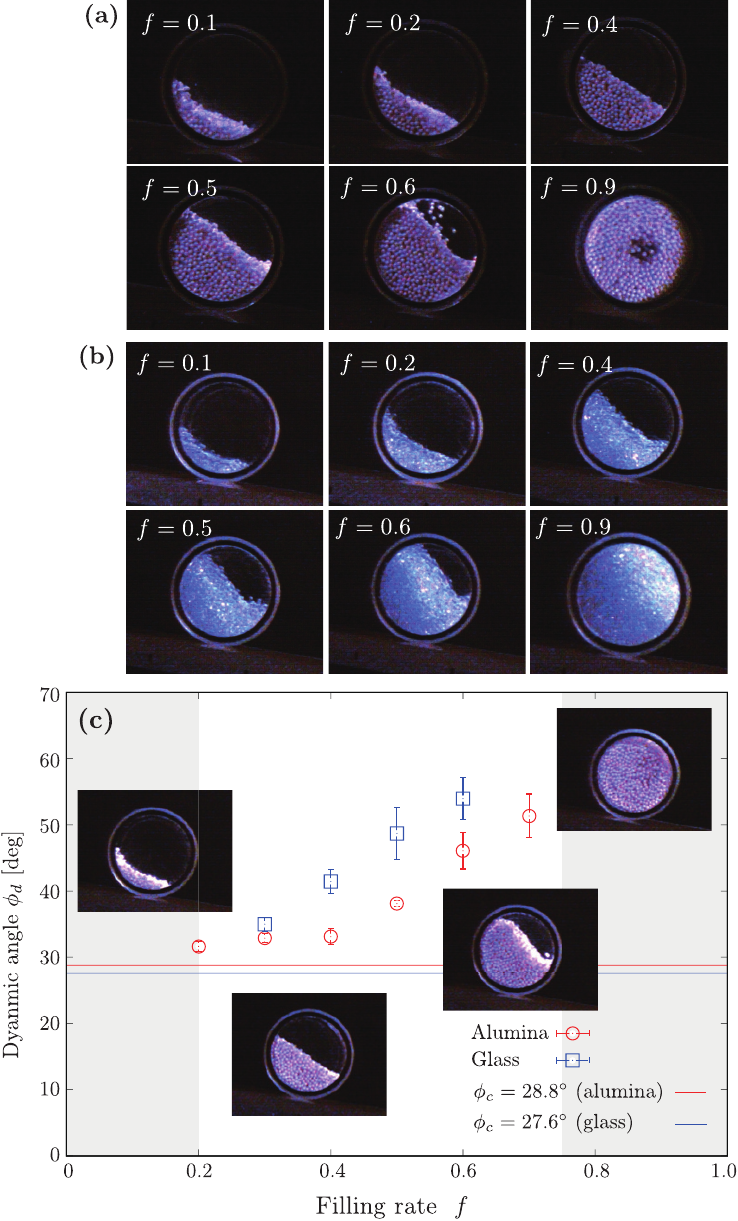}
\caption{Dynamics of granular media inside a rolling cylindrical container.
(a)-(b) Typical granular configurations captured by the high-speed camera for varying filling rates $f$ indicated: (a) alumina beads, (b) glass beads. 
(c) Dynamic angle of repose $\phi_d$ measured from the stationary granular configurations during the rolling, plotted as a function of $f$.
Insets are the typical snapshots of the alumina beads. All the data shown here is obtained for the slope angle $\alpha=12^{\circ}$.}
\label{fig:angles}
\end{center}
\end{figure}

We now consider the steady or accelerating rolling dynamics primarily represented by the red symbols in Fig.~\ref{fig:diagram}.
For $\alpha=10^{\circ}$, we used a high-speed camera to quantify the convective granular flow dynamics. 
Typical snapshots for the various $f$ are shown in Fig.~\ref{fig:angles} (a) and (b).
Remarkably, for intermediate $f$, the free granular surface is observed to be rather flat~\cite{Henein-MetTransB-1983}; thus, we define a {\it dynamic angle of repose} from the horizontal, $\phi_d$~\cite{Rajchenbach-PRL-1990, Yamane-PhysFluids-1998, Ingram-PowTech.-2005}.
As shown in Fig.~\ref{fig:angles}, $\phi_d$ increases with the filling rate $f$. 
Close to $f=1$, the angular velocities become sufficiently large so that the grains distribute almost isotropically, leaving an empty region at the center, thus making the definition of $\phi_d$ inappropriate. 
Similarly, for sufficiently low filling rates $f\ll 1$, the particles displace to the inner wall, again making the measurement of $\phi_d$ difficult.

To quantify $\phi_d$ for the intermediate $f$, we measured $\phi_d$ averaged over the section of the travel distances of 15 cm $\leq X\leq$ 35 cm from the video images, and plotted it as a function of $f$ in Fig.~\ref{fig:angles}. 
The static angles of repose, $\phi_c$, for the glass and alumina particles were measured independently, which are also plotted in Fig.~\ref{fig:angles} with the dashed lines as the references. 
As seen in Fig~\ref{fig:angles}, while the glass beads generally have larger $\phi_d$ than those of alumina beads, they both increase monotonically with $f$ to take their maximum values around $\phi_d\simeq 50^{\circ}-60^{\circ}$, reaching almost twice as large as their static counterparts, $\phi_c$. 
Beyond this regime $(f>0.8)$, the flow undergoes the transition to the ones shown in Fig.~\ref{fig:angles}, due primarily to the sufficient centrifugal forces, where $\phi_d$ is no longer well-defined. 

A major source of irreversible energy dissipation is the frictional convective granular flow inside the container. 
For $f\ll 1$, most granules move with the inner surface.
Due to the small amount of grains, the rate of energy dissipated remains quite small in this regime.
On the other hand, for $f \lesssim 1$, most grains undergo rigid-body rotations with the container, again leading to little dissipation. 
In contrast, for intermediate $f$, the strong shear flow localized near the free inclined surface becomes prominent—-see images for $f=0.4-0.5$ in Fig.~\ref{fig:angles} (a) and (b).
Such a frictional convective flow effectively dissipates kinetic energy, reducing the rolling velocity of the entire object, as observed in Fig.~\ref{fig:trajectory} (c).

\section{Summary}~\label{sec:summary}
In the final stage of this study, we became aware of the work~\cite{Wibowo-PowTech-2016} that studied a similar problem. 
Notably, the transition between rolling and stopping was analytically and experimentally investigated for different types of grains. 
Their results are fully consistent with ours, further confirming the validity of this study. 
We provided physically transparent theoretical arguments on the stability of the container on a slope and demonstrated the excellent quantitative agreement between the results of experiments and theory.
Going beyond the linearized dynamics, we established the stability phase diagram by compiling extensive experimental data from our experiments.
In addition to revealing the phase boundary separating the stopping and rolling, we also addressed the peculiar region near this boundary, in which a cylinder exhibits complex behavior, including meandering, significant slip, and off-axis stop, possibly suggesting a generic multistable nature due to the dry friction inherent to granular media inside the cylinder.
Furthermore, we generalized the concept of the angle of repose to the dynamic one, $\phi_d$, to quantify the stationally granular configurations during rolling~\cite{Rajchenbach-PRL-1990, Yamane-PhysFluids-1998, Ingram-PowTech.-2005}. 
While the experimental data clearly suggested the linear increase of $\phi_d$ as a function of the filling rate $f$, its quantitative theoretical explanation remains an intriguing future subject. 

A grain-filled rigid cylinder set on an inclined plane demonstrates a surprisingly wide variety of movements, including stop, slip, stationary, and accelerating rolling dynamics, highlighting the non-trivial coupling between translation and rotational degrees of freedom mediated by the frictional convective granular flow inside the solid body. 
Our study not only satisfies the curiosity to understand everyday phenomena (in the kitchen), but also offers a valuable avenue for harnessing and controlling granular flows in powder manufacturing technologies, as well as for innovative functionalities in robotics~\cite{Brown-PNAS-2010, Ilin-PNAS-2017, Putkaradze-Meccanica-2018, Li-Adv.Robot.-2024} and architected materials~\cite{Dierichs-Bioinspir.Biomim.-2021, Sobolev-Nature-2023}.

\begin{acknowledgments}
We thank T. Terai and M. Otsuki for sharing their preliminary numerical simulation data with us and for the ongoing collaboration.
We also thank T. Yoneda for his technical assistance with high-speed photography.
This study was supported by the Japan Society for the Promotion of Science Grant-in-Aid for Scientific Research (JSPS KAKENHI Grants No. 22H05067 to H.W.)
\end{acknowledgments}

\appendix

\section{Linear dynamics}~\label{appendix_A}
Here we present a simple theory to describe the small-amplitude damped oscillation observed in Fig.~\ref{fig:trajectory} (a) and (b).
Assuming that the mass distribution of the granular medium remains unchanged during the oscillation, our solid object can be simply modeled as a rigid cylinder with its center of mass G set off from its geometric center C by a certain distance $b$.
The dependence of $b$ on the filling rate $f$ is given in Eqs.~(\ref{eq:b}) and (\ref{eq:f}).

Since the motion is essentially planar, we use the two-dimensional plane, in which we take the tilted $x$ and $y$ axes as shown in Fig.~\ref{fig:model}.
The rotational angle is represented by $\theta(t)$ measured from the $y$ axis. 
Note that $\theta$ is taken as positive for clockwise rotation (so that the positive direction of the translation is consistent with the $x$ axis), which is opposite to the convention in the right-handed coordinate frame. 
With this setup, the position vector of G can be written as ${\bm R}_G=(X-b\sin\theta)\e_x+(-b\cos\theta)\e_y$, where $X$ denotes the travel distance along the slope.
Let A be the instantaneous contact point with the slope. 
Assuming the non-slip codnition at A, $X$ is related to $\theta$ as 
\begin{eqnarray}
 \dot{X}=R\dot{\theta},
 \label{eq:non-slip} 
\end{eqnarray}
where $\dot{()}$ represents a time derivative. 
The tangential frictional force $f_r$ acts at A as the constraining force to satisfy Eq.~(\ref{eq:non-slip}). 
The total force on the container from the slope is then expressed as ${\bm f}_A= (-f_r)\e_x+P\e_y$, where $P$ is the normal reaction force.
Newton's second law applies to the center of mass G, leading to the equations of motion for the translational and rotational degrees of freedom given by
\begin{eqnarray}
 M a_x &=& Mg\sin\alpha-f_r,
 \label{eq:eqm-x} \\
 M a_y &=& -Mg\cos\alpha+P,
 \label{eq:eqm-y}\\
 I_G\ddot{\theta} &=& - \e_z \cdot {\bm r}_{\rm GA}\times {\bm f}_A ,
 \label{eq:eqm-z}
\end{eqnarray}
where $\ddot{\bm R}_G=(a_x,a_y)$ is the acceleration vector, $M$ is the total mass of the entire object, $I_G$ is the moment of inertia about the rotational axis passing through G, and 
${\bm r}_{\rm GA} = b\sin\theta \e_x+(b\cos\theta-R)\e_y$.
Note that the negative sign in front of the r.h.s of Eq.~(\ref{eq:eqm-z}) follows from the definition of $\theta$ mentioned above.  

Using Eqs.~(\ref{eq:eqm-x}) and (\ref{eq:eqm-y}) to eliminate $f_r$ and $P$ in Eq.~(\ref{eq:eqm-z}), we finally obtain a full nonlinear time evolution equation for $\theta$ as
\begin{equation}
 \left[I_C +M(R^2-2Rb \cos\theta)\right] \ddot{\theta} = -Mgb\dot{\theta}^2 \sin\theta -\frac{dU(\theta)}{d\theta},
 \label{eq:eqm-theta}
\end{equation}
where $I_C$ is the moment of inertia about the rotational axis passing through the geometric center C of the cylinder that is related to $I_G$ as $I_C=I_G+Mb^2$, and we have defined the effective potential 
\begin{eqnarray}
 U(\theta) &=& -Mg R \theta \sin\alpha-Mg b\cos(\alpha+\theta).
 \label{eq:U}
\end{eqnarray}

For the static equilibrium configuration to exists for a given $\alpha$, the condition, $\sin\alpha<b/R$, must be satisfied; See Eq.~(\ref{eq:cond-1}).
Given this, the equilibrium cylinder angle $\theta_0$ is determined by 
\begin{equation}
 \left. \frac{dU}{d\theta}\right|_{\theta_0} = - Mg R \sin\alpha+Mg b\sin(\theta_0+\alpha) = 0.
 \label{eq:theta_0}
\end{equation}
From this mechanical balance equation, the following is obtained:
\begin{eqnarray}
 \cos\theta_0 &=& \frac{R}{b}\sin^2\alpha+\cos\alpha\sqrt{1-\frac{R^2}{b^2}\sin^2\alpha}.
 \label{eq:cos}
\end{eqnarray}

To find the eigenfrequency $\Omega$ around the stable fixed point (i.e., the static equilibrium), we assume $\theta(t)=\theta_0+\psi(t)$, where $\psi\ll 1$, and linearize Eq.~(\ref{eq:eqm-theta}) with respect to $\psi$ to find a harmonic oscillation given by $\ddot{\psi} = -\Omega^2\psi$, where 
\begin{eqnarray}
 \Omega^2 &=& \frac{Mgb\cos(\theta_0+\alpha)}{I_C+M(R^2-2bR\cos\theta_0)}.
 \label{eq:omega}
\end{eqnarray}
 
\section{Moment of inertia}~\label{appendix_B}
The moment of inertia $I_{C}$ is the sum of the contributions from the container (including the lids on both ends), $I_{\rm cyl}$, and that from the granular medium, $I_{\rm gra}$, 
that is, $I_{C}=I_{\rm cyl}+I_{\rm gra}$. 
We will examine them separately.

For $I_{\rm cyl}$, as the container has cylindrical symmetry, it is easy to find $I_{\rm cyl}=M_{0} R^2$, where $M_{0}$ is the total mass of the cylindrical container including the lids.  
To evaluate $I_{\rm gra}$, we define the angle $\beta$ to parameterize the distribution of the grains of effective mass density $\rho_g$. See Fig.~\ref{fig:experiment} (d).
We analytically calculate $I_{\rm gra}$ as follows:
\begin{eqnarray}
 I_{\rm gra} (\beta) &=& \frac{1}{8}\rho_gR^4 W\left[\beta-\frac{1}{3}\sin\beta(2+\cos\beta)\right],
 \label{eq:I_g}
\end{eqnarray}
where $W$ is the axial length of the container [Fig.~\ref{fig:experiment} (d)].
The total mass of the grains, $m_g$, can be calculated as
\begin{eqnarray}
 m_g(\beta) &=& \frac{1}{2}\rho_g (R-h)^2 W \left(\beta-\sin\beta\right),
 \label{eq:m-g}
\end{eqnarray}
where $h$ is the thickness of the side wall of the cylindrical container [Fig.~\ref{fig:experiment} (d)].
The granular filling rate, $f$, is defined as $m(\beta)=\pi (R-h)^2 W \rho_g f(\beta)$, from which we obtain Eq.~(\ref{eq:f}).

\bibliography{Rolling_Refs.bib}

\end{document}